\documentclass[letterpaper, 12 pt, onecolumn, conference]{ieeeconf}

\usepackage{amsfonts,amssymb,amsmath,epsfig,graphicx,amsmath,epsfig,epstopdf,color}

\IEEEoverridecommandlockouts
\newtheorem{theorem}{Theorem}

\newtheorem{remark}{Remark}

\newtheorem{property}{Property}

\begin{document}

\title{Position Constrained,  Adaptive Control of Robotic Manipulators without Velocity Measurements }
\author{Samet Gul, Erkan Zergeroglu and Enver Tatlicioglu\thanks{
\newline S. Gul and E. Zergeroglu are with the Department of Computer Engineering, Gebze Technical University, 41400, Gebze, Kocaeli, Turkey (Email: e.zerger@gtu.edu.tr),
\newline E. Tatlicioglu is with the Department of Electrical \& Electronics Engineering, Ege University, 35100, Bornova, Izmir Turkey  Email: enver.tatlicioglu@ege.edu.tr).
}}

\maketitle

\begin{abstract}
This work presents the design and the corresponding stability analysis of a model based, joint position tracking error constrained, adaptive output feedback  controller for robot manipulators. Specifically, provided that the initial joint position tracking error starts within a predefined region, the proposed controller algorithm ensures that the joint tracking error remains inside this region and asymptotically approaches to zero, despite the lack of joint velocity measurements and uncertainties associated with the system dynamics.  The need for the joint velocity measurements are removed via the use of a surrogate filter formulation in conjunction with  the use of desired model compensation. The stability and the convergence of the closed loop system are proved via a barrier Lyapunov function based argument. A simulation  performed on a  two-link robotic manipulator is provided in order to illustrate the feasibility and effectiveness of the proposed method. 
\end{abstract}



\section{Introduction}
Control of robotic manipulators have been a fruitful research field for the last three decades. Due to the nonlinear nature and the presence of uncertain terms in its  dynamics, the position tracking control problem for robotic devices have attracted numerous researches. The output feedback control problem, that is when only joint level position measurements are available for the control design, is among the most popular problems in the field. The reason behind this is mainly for two folds: From the theoretical perspective, designing a controller without the joint level velocity measurements is more challenging, and from practical perspective, output feedback controllers allows the manufacturers to remove one expensive sensor from each joint. Motivated by these, researchers have proposed several output feedback type controllers for robotic manipulators. To name a few Nicosia and Tomei in \cite{NicoTome90} proposed a model based observer in order to cope for the velocity measurements, in \cite{YuanStep91}, authors proposed a robust controller approach and a passivity based controller-observer formulation was proposed in \cite{BergNilm93}. Recently a model independent observer based controller was presented in \cite{ObserverOfb17}. Repetitive learning type controller approaches, where the desired trajectory needs to be periodic, were also extended to output feedback tracking control of robotic devices \cite{MerveACC15}. Filtered based approaches, where a velocity surrogate signal generated by a filter formulation is used instead of the actual velocity signal were also applied to the output feedback control of robotic manipulators \cite{BurgFilter97}. A composite adaptive scheme where the uncertain parameters of the robotic device were estimated by a combination of gradient based and least squares based estimators in conjunction with a filtered based velocity surrogate signal approach was presented in \cite{CompEZ99}. Most output feedback controllers are semi-global in nature, while global output feedback controllers with uncertain dynamics term adaptations were also presented in \cite{FangOfb20} and \cite{zerCDC00}. 


Nearly all of the previous research mentioned above, were backed up via Lyapunov type arguments for the stability and convergence of the tracking error term. One short coming of the Lyapunov based analysis is the lack of system's overshoot limitation. For robotic systems, tracking response of the system should bare nearly no overshoot. In order to reduce or minimize engineers need to tweak with the controller and estimator gains and when the controller design cannot limit the tracking error or the initial overshoot theoretically, control gain adjustments rely mainly on experience.  A solution to address this research problem relies on barrier Lyapunov functions (BLFs) \cite{jiang05}, \cite{Tee09}. In BLF based designs, bounds for each entry of the system states could be imposed \textit{a priori}. Some past research focused on applying BLF based designs to nonlinear systems of various forms \cite{jiang05}, \cite{Tee09}, \cite{Tee11}, \cite{Liu16}, \cite{Wang17}, \cite{Afflitto18}, while applying BLF type control techniques to mechatronic systems was also studied. \cite{kabzinski17} designed a BLF based systematic motion controller for servo systems. Doulgeri et al. used prescribed performance criteria for regulation control of robot manipulators in \cite{Doulgeri09}, \cite{Doulgeri10icra}, and for tracking control in \cite{Doulgeri12RAS}, \cite{Doulgeri13ROB}, \cite{Doulgeri16RAL}. \cite{Hackl12} designed a position controller based on prescribed performance criteria for robot manipulators where partial dynamic model knowledge was assumed known. \cite{Zhang18} designed a neural network based controller for robot manipulators subject to model uncertainties. Robust fixed--time control of a biped robot based on tangent BLF was presented in \cite{Rincon19}. Some task space control problems were researched in \cite{Doulgeri10iros}, \cite{Doulgeri16}, \cite{Doulgeri12}, while position/force control were addressed in \cite{Doulgeri12AUT}, \cite{Doulgeri10}. Neural network based adaptive methods in conjunction with BLFs were designed to address different control problems for Euler Lagrange systems \cite{Zhao18}, marine vessels \cite{SSGe14}, \cite{He17}, \cite{Xia19}. Review of the relevant literature reveals that when full state feedback is available several research problems were studied in conjunction with BLFs for different constraints, but when output is the only available state there are a few results obtained \cite{Rincon19}, \cite{Xia19}. 

In this work, tracking control of robot manipulators is aimed where the control problem is restricted by the unavailability of joint velocity measurements and parametric uncertainties in the mathematical model. Guaranteeing \textit{a priori} limits for each entry of joint tracking error and thus for each joint position is targeted as a secondary control task. To overcome the lack of velocity sensing, a filter based approach is preferred. In filter based approaches, via utilizing a bank of filters, a controller could be designed without needing velocity measurements. Parametric uncertainties are dealt with the design of a desired compensation based adaptive component. To restrict the entries of the joint position tracking error, a tracking error--dependent gain matrix multiplying the tracking error is proposed as part of the controller. The final form of the controller is composed of a filter vector acting as pseudo velocity tracking error in conjunction with an error--dependent gain matrix multiplying the tracking error fused with a desired compensation based adaptive component. Via introducing a BLF, in addition to proving asymptotic convergence of the tracking error to the origin, its entries are restricted to remain within user defined bounds. The simulation result obtained from a two degree of freedom robot manipulator is shown to be commensurate with the analysis.


\section{Robot Dynamic Model and Model Properties}

The mathematical model of the robot manipulators along with standard model properties are presented in this section. The dynamic model of an $n$ degree of freedom revolute joint robot manipulator has the given form \cite{lewis}
\begin{equation}
M\left(q\right)\ddot{q} + V_m\left(q ,\dot{q}\right)\dot{q} + G\left(q\right) + F_{d}\dot{q} = \tau \label{model}
\end{equation}
in which $q\left(t\right)$, $\dot{q}\left(t\right)$, $\ddot{q}\left(t\right) \in \Re^{n}$ are joint position, velocity and acceleration vectors, respectively, $M\left(q\right) \in \Re^{n\times n}$ is the inertia matrix, $V_m\left(q,\dot{q}\right) \in \Re^{n\times n}$ stands for centripetal Coriolis matrix, $G\left(q\right) \in \Re^{n}$ models gravitational effects, $F_d \in \Re^{n\times n}$ denotes diagonal viscous frictional effects, and $\tau\left(t\right) \in \Re^{n} $ is the control input torque. 

The dynamic modeling terms in \eqref{model} satisfy the following commonly utilized properties. 

\begin{property}\label{P1}
The inertia matrix is positive definite and symmetric and satisfies \cite{lewis} 
\begin{equation}
m_1 I_{n} \leq M \left(q\right) \leq m_2 I_{n} \label{prop1}
\end{equation}
in which $m_1$ and $m_2$ are known positive bounding constants, and $I_n$ stands for $n$--by--$n$ identity matrix.
\end{property}

\begin{property}\label{P2}
The inertia and centripetal Coriolis matrices satisfy \cite{lewis}
\begin{equation}
\xi^T \left(\dot{M}-2V_m\right) \xi = 0 \quad \forall \xi \in \Re^{n}. \label{prop2}
\end{equation}
\end{property}

\begin{property}\label{P3}
The centripetal Coriolis matrix satisfies \cite{lewis}
\begin{equation}
V_m\left(\xi ,\nu\right)\eta = V_m\left(\xi ,\eta\right) \nu \quad \forall \xi, \nu, \eta \in \Re^{n}. \label{prop3}
\end{equation}
\end{property}

\begin{property}\label{prop4}
Following bounds can be proven to be satisfied for the dynamic modeling terms in \eqref{model} \cite{lewis}
\begin{eqnarray}
\Vert M\left( \xi \right) - M\left( \nu \right) \Vert_{i\infty} & \leq & \zeta_{m1} \Vert \xi - \nu \Vert \label{prop4a} \\
\Vert V_m\left( \xi ,\nu \right) \Vert_{i\infty} & \leq & \zeta_{c1} \Vert \nu \Vert \label{prop4b} \\
\Vert V_m\left( \xi ,\nu \right) - V_m\left( \eta , \nu\right) \Vert_{i\infty} &\leq & \zeta_{c2} \Vert \xi - \eta \Vert \Vert \nu \Vert \label{prop4c} \\
\Vert G\left( \xi \right) - G\left( \nu\right) \Vert & \leq & \zeta_{g} \Vert \xi - \nu \Vert \label{prop4d}
\end{eqnarray}
$\forall \xi $, $\nu$, $\eta \in \Re^{n}$ with $\zeta_{m1}$, $\zeta_{c1}$, $\zeta_{c2}$, $\zeta_{g}\in \Re$ being known, positive bounding constants.
\end{property}

\begin{property}\label{P5}
The left hand side of \eqref{model} can be reconfigured to be written in linearly parameterized form as
\begin{equation}
Y\left(q,\dot{q},\ddot{q}\right)\theta = M\left( q\right)\ddot{q} + V_m\left( q,\dot{q}\right)\dot{q} + G\left( q\right) + F_{d}\dot{q} \label{prop5}
\end{equation}
with $Y\left(q,\dot{q},\ddot{q}\right)\in \Re^{n\times p}$ being a regression matrix and $\theta \in \Re^{p}$ containing constant model parameters that depends on physical properties of the robot manipulator. A desired form of the structure in \eqref{prop5} can be written as
\begin{equation}
Y_d \theta = M\left( q_{d}\right)\ddot{q}_{d} + V_m\left( q_{d},\dot{q}_{d}\right)\dot{q}_{d} + G\left( q_{d}\right) + F_{d}\dot{q}_{d} \label{prop5d}
\end{equation}
in which $Y_d\left(q_{d},\dot{q}_{d},\ddot{q}_{d}\right)\in \Re^{n\times p}$ is the desired version of the regression matrix that is a function of the desired joint position, velocity and acceleration vectors, denoted respectively with $q_{d}\left( t\right)$, $\dot{q}_{d}\left( t\right)$, $\ddot{q}_{d}\left( t\right) \in \Re^{n}$.
\end{property}

\section{Control Objective, Error System Development and Design}

In this section, the control objective along with the restrictions will be presented first, the error system will then be introduced, and the control input torque will be designed. Ensuring tracking of a sufficiently smooth\footnote{The desired joint position trajectory along with its first two time derivatives are to be designed as bounded functions of time.} desired joint position vector $q_{d}\left( t\right)$ via designing the control input torque $\tau \left( t\right)$ is the primary control objective. In addition to the tracking objective, each joint's position is required to remain within a predefined neighborhood $\Delta_{i}$ of its desired joint position. Guaranteeing stability of the closed loop system via ensuring boundedness of all the signals is also essential. The control design is restricted by the unavailability of joint velocity measurements (\textit{i.e.}, only joint position $q\left( t\right) $ is available for using in control) and the mathematical model in \eqref{model} is subject to structured uncertainties in the sense that $\theta$ in Property \ref{P5} is unknown.

To quantify the main control objective, the joint position tracking error, shown with $e\left( t\right) \in \Re^{n}$, is introduced as
\begin{equation}
e \triangleq q_{d} - q \label{e}
\end{equation}
and in view of the above definition, the secondary control objective can be formulated as
\begin{equation}
\vert e_{i} \left(t\right) \vert < \Delta_{i} \quad \forall t>0 \text{ , } i \in \lbrace 1, \cdots, n \rbrace . \label{obj}
\end{equation}

Since joint velocity measurements are not available for using in the control design, a filter based approach is to be preferred. Specifically, $e_f\left(t\right) \in \Re^n$, which will be used instead of the actual velocity error, is designed as the output of the following filter
\begin{equation}
e_{f} = -ke + w \label{ef} 
\end{equation}
in which $k$ is a positive gain and $w\left(t\right) \in \Re^n$ is updated according to
\begin{equation} 
\dot{w} = -\left(k+1 \right)e_{f} - k e + K_{e}e \quad w\left(0\right) = k e\left(0\right) \label{wdot}
\end{equation}
where $K_{e} \left(e\right) \in \Re^{n\times n}$ is an error--dependent gain matrix designed as 
\begin{equation}
K_{e} = \text{diag} \left\{ \frac{K_{i}}{\Delta_{i}^{2} - e_{i}^{2}} \right\} \label{Keln}
\end{equation}
with $K_{i}$ $i \in \lbrace 1, \cdots ,n \rbrace$ being constant gains. To obtain the dynamics for $e_f\left(t\right)$, time derivative of \eqref{ef} is taken to which \eqref{wdot} is substituted into to deduce
\begin{equation}
\dot{e}_{f} = - e_{f} - k\eta + K_{e}e \label{efdot}
\end{equation}
with $\eta\left(t\right) \in \Re^n$ being another filter term defined as
\begin{equation}
\eta \triangleq \dot{e} + e + e_{f} . \label{eta}
\end{equation}
Due to the time derivative of the tracking error in its definition, $\eta\left(t\right)$ is not an available quantity. The dynamics for position tracking error is obtained after rearranging \eqref{eta} as
\begin{equation}
\dot{e} = - e - e_{f} + \eta . \label{edot}
\end{equation}
To obtain the dynamics of $\eta\left(t\right)$, the time derivative of \eqref{eta} is pre--multiplied with $M\left(q\right)$ to reach
\begin{eqnarray}
M\left(q\right)\dot{\eta} &=& M\left(q\right) \left(\ddot{q}_{d} + \dot{e} + \dot{e}_{f} \right) \nonumber \\
                          &+& V_m\left(q ,\dot{q}\right) \dot{q} + G\left(q\right) + F_{d}\dot{q} - \tau \nonumber 
\end{eqnarray}
where \eqref{model} was substituted into. Adding and subtracting the desired dynamics in Property \ref{P5} to the right hand side of the above expression and making use of \eqref{efdot} and \eqref{edot} yields
\begin{equation}
M\left(q\right) \dot{\eta} = - V_m\left(q,\dot{q}\right)\eta - k M\left(q\right)\eta + Y_d \theta + \chi - \tau \label{Metadot}
\end{equation}
in which $\chi\left(e,e_f,\eta,t\right) \in \Re^n$ has the following structure
\begin{eqnarray}
\chi & \triangleq& M\left(q\right) \left( \ddot{q}_d + \eta - 2 e_f + K_e e - e \right) \nonumber \\
     &+& V_m\left(q,\dot{q}\right)\left(\dot{q}_d + e_f + e\right) \nonumber \\
     &+& G\left(q\right) + F_d\dot{q} - Y_d\theta . \label{chi}
\end{eqnarray}
Making use of the boundedness properties of the dynamic modeling terms, an upper bound can be deduced for $\chi$ as
\begin{equation}
\Vert \chi \Vert \leq \zeta_1 \Vert x \Vert + \zeta_2 \Vert x \Vert^2 \label{chiBound}
\end{equation}
in which $\zeta_1$ and $\zeta_2$ are known positive bounding constants and $x\left(t\right) \in \Re^{3n}$ is the combined error vector that has the form 
\begin{equation}
x \triangleq \left[ \eta^T \quad e_f^T \quad e^T \right]^T . \label{xdef}
\end{equation}

The control input torque is designed as
\begin{equation}
\tau = Y_{d}\hat{\theta} + K_e e - k e_f \label{tau}
\end{equation}
where $\hat{\theta}\left(t\right) \in \Re^{p}$ is the adaptive estimate of the uncertain parameter vector $\theta$ that was introduced in Property \ref{P5} and its update law is designed as 
\begin{eqnarray}
\hat{\theta} &=& \Gamma \int_{0}^{t} Y_d^T \left(\sigma\right) \left(e_f\left(\sigma\right) + e\left(\sigma\right)\right) d\sigma \nonumber \\
             &+& \Gamma Y_d^T e - \Gamma \int_{0}^{t} \frac{d\lbrace Y_d^T\left(\sigma\right) \rbrace}{d\sigma} e\left(\sigma\right) d\sigma \label{hattheta}
\end{eqnarray}
in which $\Gamma \in \Re^{p\times p}$ is constant, diagonal, positive definite adaptation gain matrix. In view of \eqref{eta}, the time derivative of \eqref{hattheta} yields
\begin{equation}
\dot{\hat{\theta}} = \Gamma Y_d^T \eta . \label{uRule}
\end{equation}
Substituting the control input torque in \eqref{tau} into \eqref{Metadot} gives
\begin{eqnarray}
M\left(q\right)\dot{\eta} &=& - V_m\left(q,\dot{q}\right)\eta - k M\left(q\right)\eta \nonumber \\
                          &-& K_e e + k e_f + Y_d\tilde{\theta} + \chi \label{MetadotCL}
\end{eqnarray}
with $\tilde{\theta}\left(t\right) \in \Re^{p}$ representing the parameter estimation error as
\begin{equation}
\tilde{\theta} \triangleq \theta - \hat{\theta} . \label{tildetheta}
\end{equation}

\section{Analysis}

The stability analysis for the filter based output feedback control strategy will be investigated in this section. Following theorem frames the stability analysis. 

\begin{theorem} \label{Thm}
The control input torque in \eqref{tau} in conjunction with the parameter update law in \eqref{hattheta} and the filter design of \eqref{ef} and \eqref{wdot} ensures boundedness of all the signals under the closed loop system and semi--global asymptotic tracking provided that $\min\lbrace K_i \rbrace \geq \max \lbrace \Delta_i^2 \rbrace$ is satisfied and the control gain $k$ is designed as
\begin{equation}
k = \frac{1}{m_1} \left(1 + \zeta_1^2 k_n + \zeta_2^2 k_n\right) \label{k}
\end{equation}
with $k_n$ being a nonlinear damping gain designed to satisfy
\begin{equation}
k_n > 1 + \frac{\lambda_2}{\lambda_1} \Vert z\left(0\right) \Vert^2 \label{kn}
\end{equation}
in which $z\left(t\right) \triangleq \left[x^T \quad \tilde{\theta}^T \right]^T \in \Re^{3n+p}$, and $\lambda_1$ and $\lambda_1$ are positive constants defined as\footnote{In \eqref{lambdas}, the notations $\lambda_{\min}\lbrace \cdot \rbrace$ and $\lambda_{\max}\lbrace \cdot \rbrace$ denote, respectively, minimum and maximum eigenvalue of a matrix.}
\begin{eqnarray*}
&&\lambda_1 \triangleq \frac{1}{2} \min \lbrace m_1, 1, \frac{\min \lbrace K_i \rbrace}{ \max \lbrace \Delta_{i}^{2}\rbrace },  \lambda_{\min} \lbrace \Gamma^{-1} \rbrace \rbrace , \\
&&\lambda_2 \triangleq \frac{1}{2} \max \lbrace m_2, 1, \max \lbrace K_e \rbrace,  \lambda_{\max} \lbrace \Gamma^{-1} \rbrace \rbrace . \label{lambdas}
\end{eqnarray*}

\end{theorem}

The proof is initiated via the definition of a barrier Lyapunov function denoted with $V\left(\eta,e_f,e,\tilde{\theta}\right)$ as
\begin{eqnarray}
V &\triangleq & \frac{1}{2}\eta^{T}M\left( q\right)\eta + \frac{1}{2}e_f^{T}e_f \nonumber \\
  &+& \sum_{i=1}^{n}\frac{K_{i}}{2}\ln\left( \frac{\Delta_{i}^{2}}{\Delta_{i}^{2}-e_{i}^{2}}\right) + \frac{1}{2}\tilde{\theta}^{T}\Gamma^{-1}\tilde{\theta} \label{Vl}
\end{eqnarray}
which can be bounded via
\begin{equation}
\lambda_1 \Vert x \Vert^2 \leq \lambda_1 \Vert z \Vert^2 \leq V \leq \lambda_2 \Vert z \Vert^2 . \label{VlBound}
\end{equation}
Provided that the initial error values of all joints satisfy $\vert e_{i}\left( 0\right) \vert < \Delta_{i}$, $V\left(\eta,e_f,e,\tilde{\theta}\right)$ is positive definite and radially unbounded thus qualifies as a barrier Lyapunov function. 

Taking the time derivative of $V$ gives 
\begin{eqnarray}
\dot{V} &=& \frac{1}{2}\eta^{T}\dot{M}\left(q\right)\eta + \eta^T M\left(q\right)\dot{\eta} + e_f^{T} \dot{e}_f \nonumber \\
        &+& \sum_{i=1}^{n} \frac{e_i K_{i} \dot{e}_i}{\Delta_{i}^2 - e_i^2} + \tilde{\theta}^{T} \Gamma^{-1}\dot{\tilde{\theta}} \label{dotVlBound1}
\end{eqnarray}
and after noting the structure of \eqref{Keln}, the summation can be reformulated as
\begin{equation}
\sum_{i=1}^{n} \frac{e_i K_{i} \dot{e}_i}{\Delta_{i}^2 - e_i^2} = e^T K_e \dot{e} . \label{Keln2}
\end{equation}
Substituting \eqref{efdot}, \eqref{edot} in view of \eqref{Keln2}, \eqref{uRule} in view of time derivative of \eqref{tildetheta}, \eqref{MetadotCL} into $\dot{V}$ and  making use of the skew--symmetry of the robot dynamics in Property \ref{P2} gives
\begin{eqnarray}
\dot{V} &=& \eta^T \left[ - k M\left(q\right)\eta - K_e e + k e_f + Y_d\tilde{\theta} + \chi \right] \nonumber \\
       &+& e_f^{T} \left(- e_{f} - k\eta + K_{e}e\right) \nonumber \\
       &+& e^T K_e \left( - e - e_{f} + \eta\right) - \tilde{\theta}^{T} Y_d^T\eta\label{dotVlBound2}
\end{eqnarray}
which after  canceling common terms, deduces to
\begin{equation}
\dot{V} = - e_f^{T} e_{f} - e^T K_e e - k \eta^T M\left(q\right)\eta + \eta^T\chi . \label{dotVlBound3}
\end{equation}
In view of Property \ref{P1}, 
\begin{equation}
- k \eta^T M\left(q\right)\eta \leq - k m_1 \Vert \eta \Vert^2 \label{Bound1}
\end{equation}
and from the structure of \eqref{Keln} and the condition of $\min\lbrace K_i \rbrace \geq \max \lbrace \Delta_i^2 \rbrace$,
\begin{equation}
- e^T K_e e \leq -\frac{\min\lbrace K_i \rbrace}{\max\lbrace \Delta_{i}^2 \rbrace} \Vert e \Vert^2 \leq - \Vert e \Vert^2 \label{Bound2}
\end{equation}
are satisfied. After substituting \eqref{k} and the bounds of \eqref{chiBound}, \eqref{Bound1} and \eqref{Bound2}, the right hand side of \eqref{dotVlBound3} reaches
\begin{eqnarray}
\dot{V} &\leq &- \Vert x \Vert^2 + \left[\xi_1 \Vert \eta \Vert \Vert x \Vert - \xi_1^2 k_n \Vert \eta \Vert^{2}\right] \nonumber \\
        & + & \left[\xi_2 \Vert \eta \Vert \Vert x \Vert^2 - \xi_2^2 k_n \Vert \eta \Vert^{2}\right] \label{dotVlBound4}
\end{eqnarray}
to which applying the nonlinear damping argument in \cite{Kokotovic92} yields
\begin{equation}
\dot{V} \leq - \left[1 - \frac{1}{4k_n} \left(1+\Vert x \Vert^2\right)\right] \Vert x \Vert^2 . \label{dotVlBound5}
\end{equation}
The sign of the upper bound on $\dot{V}$ is determined by the bracketed term of the preceding inequality and when it is positive, negative semi--definiteness of $\dot{V}$ is ensured. Mathematically, 
\begin{equation}
1 - \frac{1}{4k_n} \left(1+\Vert x \Vert^2\right) > 0 \label{dotVlBound5a}
\end{equation} 
is required for $\dot{V}$ to be negative semi--definite. In view of \eqref{VlBound}, a more conservative bound can be derived as
\begin{equation}
1-\frac{1}{4k_n}\left(1+\frac{V}{\lambda_1}\right) > 0 \label{dotVlBound5b}
\end{equation} 
which yields 
\begin{equation}
\dot{V} \leq - \beta \Vert x \Vert^2 \text{ for } k_n > \frac{1}{4}\left(1+\frac{V}{\lambda_1}\right) \label{dotVlBound6}
\end{equation}
where $0<\beta<1$. From the structures of \eqref{Vl} and \eqref{dotVlBound6}, it is clear that $V$ is non--increasing, which allows us to have
\begin{equation}
\dot{V} \leq - \beta \Vert x \Vert^2 \text{ for } k_n > \frac{1}{4}\left(1+\frac{V\left(0\right)}{\lambda_1}\right) \label{dotVlBound7}
\end{equation}
and utilizing \eqref{VlBound} gives 
\begin{equation}
\dot{V} \leq -\beta \Vert x \Vert^2 \text{ for } k_n > \frac{1}{4}\left(1+\frac{\lambda_2}{\lambda_1}\Vert z\left(0\right) \Vert^2\right) .\label{dotVlBound8}
\end{equation}

The direct implication of \eqref{Vl} and \eqref{dotVlBound8} is that $z\left(t\right)$ is bounded and thus $e\left(t\right)$, $e_f\left(t\right)$, $\eta\left(t\right)$, $\tilde{\theta}\left(t\right) \in \mathcal{L}_{\infty}$. The boundedness of $e\left(t\right)$ along with the desired joint trajectory vector being bounded implies that $q\left(t\right) \in \mathcal{L}_{\infty}$. After using the above boundedness statements with \eqref{efdot} and \eqref{eta}, $\dot{e}_f\left(t\right)$, $\dot{e}\left(t\right) \in \mathcal{L}_{\infty}$ can be proven; hence, $\dot{q}\left(t\right)$ is bounded since $\dot{q}_d\left(t\right)$ is bounded. Since $\tilde{\theta}\left(t\right)$ is bounded and $\theta$ is constant, then $\hat{\theta}\left(t\right) \in \mathcal{L}_\infty$. From \eqref{ef} and \eqref{wdot}, it is clear $w\left(t\right)$, $\dot{w}\left(t\right)$ are bounded. In view of the above boundedness statements, from its design in \eqref{tau}, $\tau\left(t\right)$ is bounded. From \eqref{MetadotCL}, it can be proven that $\dot{\eta}\left(t\right)\in \mathcal{L}_\infty$. All the remaining signals can be guaranteed to be bounded via the above boundedness results. After integrating \eqref{dotVlBound8} in time from initial time $0$ to $+\infty$, $x\left(t\right) \in \mathcal{L}_2$ is obtained. Since $x\left(t\right) \in \mathcal{L}_{2} \cap \mathcal{L}_\infty$ and $\dot{x}\left(t\right) \in \mathcal{L}_\infty$, in view of Barbalat's Lemma \cite{khalil}, $x \to 0$ as $t \to +\infty$ is proven and thus asymptotic tracking is ensured.

 We would like to note that, due to the gain condition \eqref{kn}, as one of the controller gains rely on the initial conditions of the system (semi--global stability), one might conclude that in order to extend the stability region the value of $k_n$ would selected to be high. However as the initial value of the error signal is required to start inside a predefined region, the maximum value of $k_n$ can be calculated a priori and does not necessarily have high values.

\begin{remark}
It is highlighted that instead of the $K_e$ design in \eqref{Keln} below tangent function based design could have also been utilized
\begin{equation}
K_{e} = \text{diag} \left\{ 1+\tan^{2}\left( \frac{\pi}{2} \frac{e_{i}^{2}}{\Delta_{i}^{2}}\right) \right\} \label{Ketr}
\end{equation}
which would have resulted in a similar stability result after changing the third term in \eqref{Vl} to
\begin{equation}
\sum_{i=1}^{n}\frac{\Delta_i^2}{\pi} \tan\left( \frac{\pi}{2} \frac{e_{i}^{2}}{\Delta_i^2}\right) . \label{Vt}
\end{equation}
\end{remark}

\section{Simulation}\label{expResults}
A simulation with the position constraint of 7 degrees is performed in order to see the effect of the $K_e$ term defined in equation \eqref{Keln} of the control torque input. In order to illustrate the feasibility and performance of the proposed position constrained adaptive output feedback controller, a simulation was preformed.
In the simulation, the following desired joint trajectories are applied
\begin{equation}
q_d \left(t\right) = \left[ \begin{array}{c} 0.7 \sin\left(t\right)\left(1-e^{-0.3t^3}\right) \\ 1.2\sin\left(t\right)\left(1-e^{-0.3t^3}\right) \end{array} \right] \left[\text{rad} \right] \label{expdesiredtraj}
\end{equation}
where the exponential term in the above equation is applied to give a smooth start to the system. 

The links of the robot arm were started with $2.9$ degrees initial position error, the initial parameter estimates were set to zero and the torque outputs were saturated at  $\pm 10$ newton-meters. The selected control and adaptation gains were as follows  

\[
\begin{array}{c}
~~k_{e}=diag\{%
\begin{array}{cc}
2 & 2%
\end{array}%
\},~~~k=diag\{%
\begin{array}{cc}
80 & 20%
\end{array}%
\} \\ 
\text{and} \\ 
\Gamma =\left[ 
\begin{array}{ccccc}
50 & 0.5 & 1 & 80 & 2.5%
\end{array}%
\right] ^{T}%
\end{array}%
\]

The results of the simulation are presented in Figure 1, where the top two sub--figures are the tracking error performances, the sub--figure in the middle is the parameter estimations and the bottom two sub--figures are the corresponding control torque inputs. As can be seen from Figure 1, after around 30 seconds. The parameter estimates converge to some values and the tracking error performances for link 1 and 2 converge to values below $\pm 0.05$ degrees. We can conclude that the tracking performance of the proposed controller is quite satisfactory.

\begin{figure*}
\setlength{\fboxsep}{0pt}%
\setlength{\fboxrule}{0pt}%
\begin{center}
\includegraphics[width=0.53\linewidth]{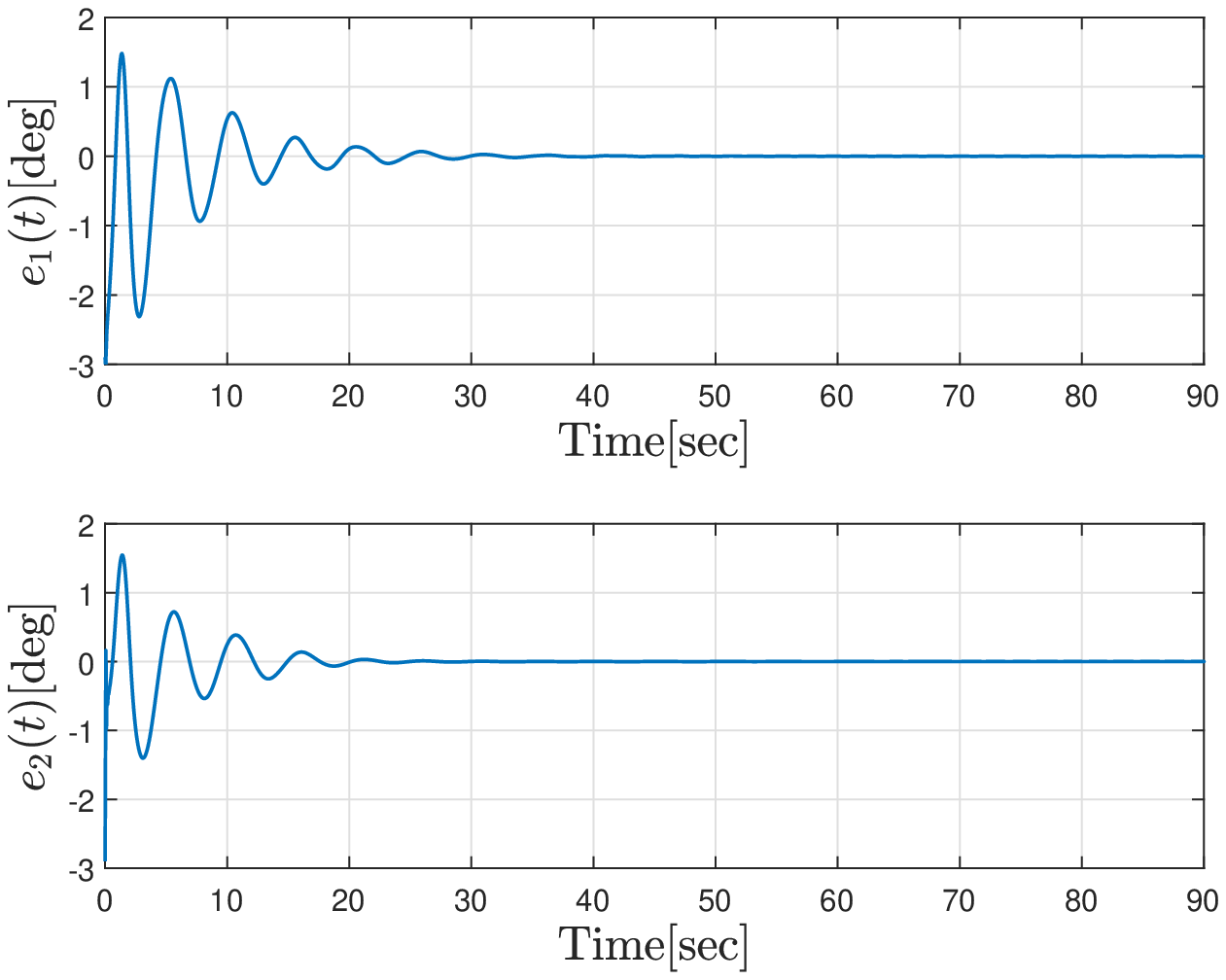}\\
\includegraphics[width=0.53\linewidth]{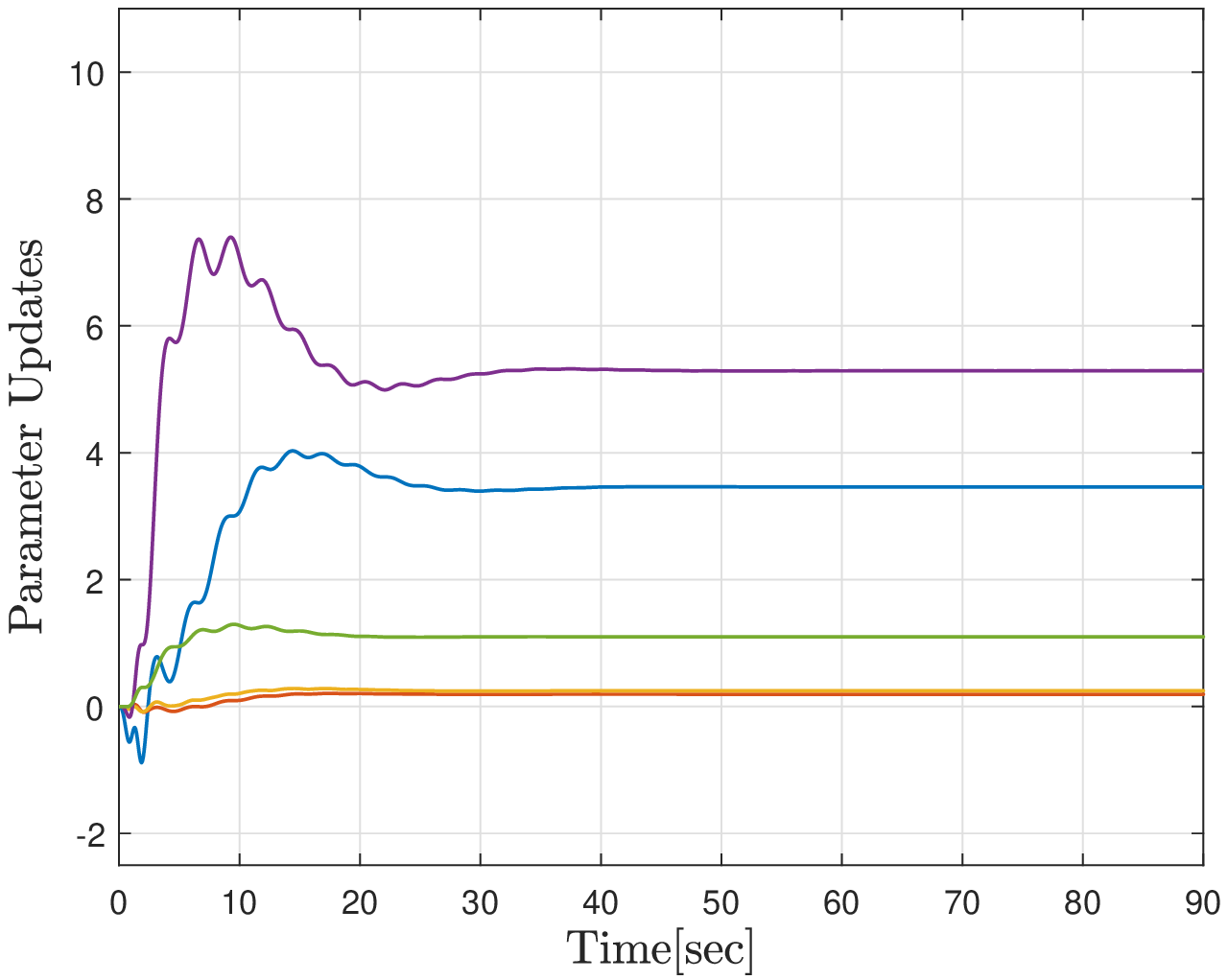}\\
\includegraphics[width=0.53\linewidth]{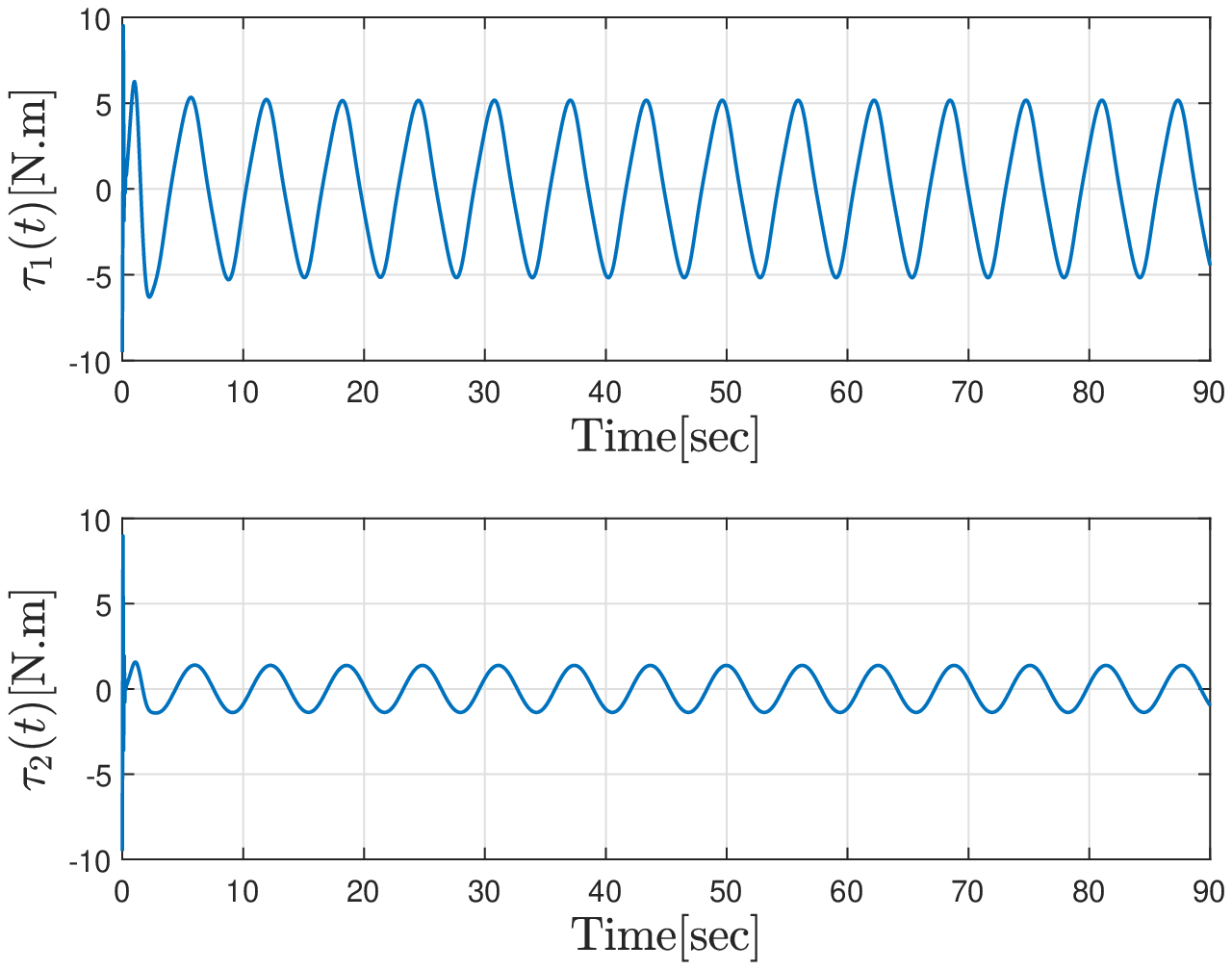} 
\end{center}
\caption{The simulation results of the output feedback controller with 7 degree constraint}
\end{figure*}

\section{Conclusions}
In this work, the design of a joint position tracking error constrained output feedback controller for robot manipulators with uncertain dynamical parameters have been presented. The proposed controller ensures that the joint tracking error signals of each link stay inside a predefined bound and eventually converge to zero, despite the lack of velocity measurements and presence of dynamical uncertainties. A simulation is also presented to illustrate the effectiveness and performance of the proposed output feedback controller. Future work will concentrate on extending the results to output feedback learning type controllers  and task space position constrained control of robotic manipulators.

%

\end{document}